\documentclass[twocolumn]{aastex62}

\usepackage{gensymb}

\graphicspath{{./}{figures/}}

\received{January 1, 2018}
\revised{January 7, 2018}
\accepted{\today}
\submitjournal{ApJ}

\shorttitle{The TRGB distance to NGC1052-DF4}
\shortauthors{Monelli \& Trujillo}

\begin{document}

\title{The TRGB distance to the second galaxy ``missing dark matter"\\ Evidence for two groups of galaxies at 13.5 and 19 Mpc in the line of sight of NGC1052}

\correspondingauthor{Matteo Monelli, Ignacio Trujillo}
\email{monelli@iac.es, trujillo@iac.es}

\author[0000-0001-5292-6380 ]{Matteo Monelli}
\affil{Instituto de Astrof{\'i}sica de Canarias (IAC), La Laguna, 38205, Spain}
\affil{Departamento de Astrof{\'i}sica, Universidad de La Laguna (ULL), E-38200, La Laguna, Spain}

\author[0000-0001-8647-2874]{Ignacio Trujillo}
\affil{Instituto de Astrof{\'i}sica de Canarias (IAC), La Laguna, 38205, Spain}
\affil{Departamento de Astrof{\'i}sica, Universidad de La Laguna (ULL), E-38200, La Laguna, Spain}



\begin{abstract}

A second galaxy ``missing dark matter" (NGC1052-DF4) has been recently reported. Here we show, using the location of the Tip of the Red Giant Branch (TRGB), that the distance to this galaxy is 14.2$\pm$0.7 Mpc. This locates the galaxy 6 Mpc closer than previously determined.  We also analyse the distances to the brightest galaxies in the field-of-view (FOV) of NGC1052. We find this field is populated by two groups of galaxies in projection: one dominated by NGC1052 and NGC1047 at $\sim$19 Mpc and another group containing NGC1042 and NGC1035 (as well as [KKS2000]04 and  NGC1052-DF4) at $\sim$13.5 Mpc. At a distance of 13.5 Mpc the globular clusters of NGC1052-DF4 have the same properties than globular clusters in the Milky Way and other dwarf galaxies.  

\end{abstract}



\section{Introduction} \label{sec:intro}

The determination of the physical properties of the astronomical objects relies on our ability to have an accurate determination of their distances. However, measuring reliable distances to most of the astronomical sources remains as one of the most difficult enterprise in our science. Recently, \citet{2018Natur.555..629V} and \citet{2019ApJ...874L...5V} have reported the presence of two extraordinary low surface brightness galaxies ([KKS2000]04 and NGC1052-DF4) in the field-of-view of the bright galaxy NGC1052. These two galaxies are proposed to have a dark matter content compatible with zero. Such claim is based on the assumption that both galaxies are at a distance of 20 Mpc.

The distance to both dwarf galaxies was measured using the surface brightness fluctuation (SBF) technique. In particular, \citet{2019ApJ...874L...5V} used the  calibration provided by \citet{2010ApJ...724..657B} to derive a distance of $\sim$20 Mpc. Unfortunately, the reliability of using a calibration which is established for red and massive galaxies to determine the distance to bluer and less massive galaxies is not at all obvious.  In fact, \citet{2018RNAAS...2c.146B} warn about the use of such a calibration in a range where it has not been explored: ``The SBF method is not well-tested at these colors and low stellar densities...". A warning that has been vindicated by recent analysis of the SBF methodology  exploring low mass galaxies \citep[see e.g.][]{2018ApJ...856..126C,2019arXiv190107575C} which directly shows that the extrapolation of \citet{2010ApJ...724..657B} is inaccurate\footnote{The extrapolation of the relation provided by \citet{2010ApJ...724..657B} is offset by around 0.55 mag for these bluer and low mass galaxies. This corresponds to an artificial increase in  the distance of $\sim$4.5 Mpc.} to measure the distance to these types of galaxies. Fortunately, there are alternative ways to measure the distance to these diffuse systems.

When galaxies are nearby enough so their stellar populations can be resolved into stars, one of the most precise determination of their distances can be obtained by  the location of the TRGB in the colour-magnitude diagram (CMD) of their stars. In a recent paper \citep{2019MNRAS.486.1192T}, we show that the location of the TRGB of the CMD of the galaxy [KKS2000]04 (popularised as NGC1052-DF2) gives a distance of 13.4$\pm$1.1 Mpc\footnote{Together with another four independent distance indicators the final distance was established on 13.0$\pm$0.4 Mpc.}. This result was shown to be robust against crowding effects as demonstrated by the fact that the distance to the object was independent to the use of stars in the innermost central region or in the periphery of the object. In this Letter, we analyze the CMD of the other dwarf ``lacking dark matter" galaxy: NGC1052-DF4. We will show that the location of the TRGB of this galaxy puts this object at a distance of 14.2$\pm$0.7 Mpc.
Finally, we also make an analysis of the distances to the brightest galaxies in the FOV of NGC1052. We find tantalising evidence that there are two groups of objects located in the same FOV: one dominated by NGC1052 at $\sim$19 Mpc and another one centred on NGC1042 at $\sim$13.5 Mpc.

\section{Data}

Hubble Space Telescope (HST) imaging of the NGC1052-DF4 galaxy explored in this work was taken as part of the programme  GO-14644  (PI: van Dokkum) and consists of two Advanced Camera for Survey (ACS) orbits for each object: one  in \textsl{F606W} (\textsl{V$_{606}$}; 2180s) and one in \textsl{F814W} (\textsl{I$_{814}$}; 2320s). A first analysis of this data was conducted by \citet{2018ApJ...868...96C}. Data were retrieved from 
the MAST  archive webpage\footnote{\url{https://mast.stsci.edu/portal/Mashup/Clients/Mast/Portal.html}}. Each orbit is split into four different images whose individual exposures are of 545s (\textsl{F606W}) and 580s (\textsl{F814W}). The pixel size is 0.05\arcsec.

The ACS exposures we used were the ones produced by the standard STScI pipeline: i.e. bias- and dark-current-subtracted, flat-fielded and with the CTE correction applied. This  produces the calibrated "flc" images. To create the CMDs of all the diffuse galaxies we follow the prescriptions given by  \citet{2010ApJ...720.1225M}. The photometry was made on the individual $flc$ images using the \texttt{DAOPHOT/ALLFRAME} set of programs \citep{stetson87,stetson94}. We extensively masked the background galaxies in our images to remove potential contamination in our catalogue by artificial deblending of these sources. Our code performs a simultaneous data reduction of the images of a given target, assuming individual Point Spread Functions (PSFs) once an input list of stellar objects is provided. The final list of point-sources was generated on the stacked median image obtained by registering and co-adding the eight individual available frames (four per filter) of each galaxy. The source detection algorithm was iterated twice. The photometry of the stars is given in the Vega system, adopting the zero-points appropriate for the observations date ($zp_{F606W}$ = 26.404, $zp_{F814W}$ = 25.516)\footnote{\url{https://acszeropoints.stsci.edu/}}. The photometric catalogue was finally cleaned using quality cuts. We used the {\itshape sharpness} parameter provided by \texttt{DAOPHOT}, rejecting sources with $|sha| >$ 0.1. The use of the {\itshape sharpness} parameter for effectively cleaning the sample of false detections (i.e. galaxies, unrecognised doubles, bad pixels or cosmic rays) has been routinely adopted in the literature \citep[see e.g.][]{2007ApJ...667L.151A,2017AJ....153..199T}. 

\section{The TRGB distance to NGC1052-DF4}

An accurate estimate of the distance to the galaxies can be obtained using the TRGB when the stellar populations  are resolved \citep{1993ApJ...417..553L}. The TRGB marks the end of the red giant branch phase, producing a  well-defined discontinuity in the luminosity function of the stars, which can be easily identified from photometric data. 
This method has two strong advantages: first, the TRGB is an intrinsically bright feature (\textsl{M$_{I}$}$\sim$ -4) and, second, the luminosity of the TRGB in the \textsl{I/F814W} filters has a very mild dependence on the age and the metallicity \citep{1997MNRAS.289..406S}. For such reasons, the method can be reliably used  beyond 10 Mpc. We took the calibration by \citet{2007ApJ...661..815R} 

\begin{equation}
M_{F814W}^{ACS} = -4.06+ 0.20\times[(F606W-F814W) - 1.23] 
\label{eq:trgb}
\end{equation}

which accounts for the mild dependency on the metallicity by taking into account a color term.  We applied a color correction to the \textsl{F814W} photometry as suggested by \citet{mcquinn17a}. This results in a steeper RGB and therefore a better defined TRGB.

    \begin{figure*}
  \includegraphics[width=0.85\textwidth]{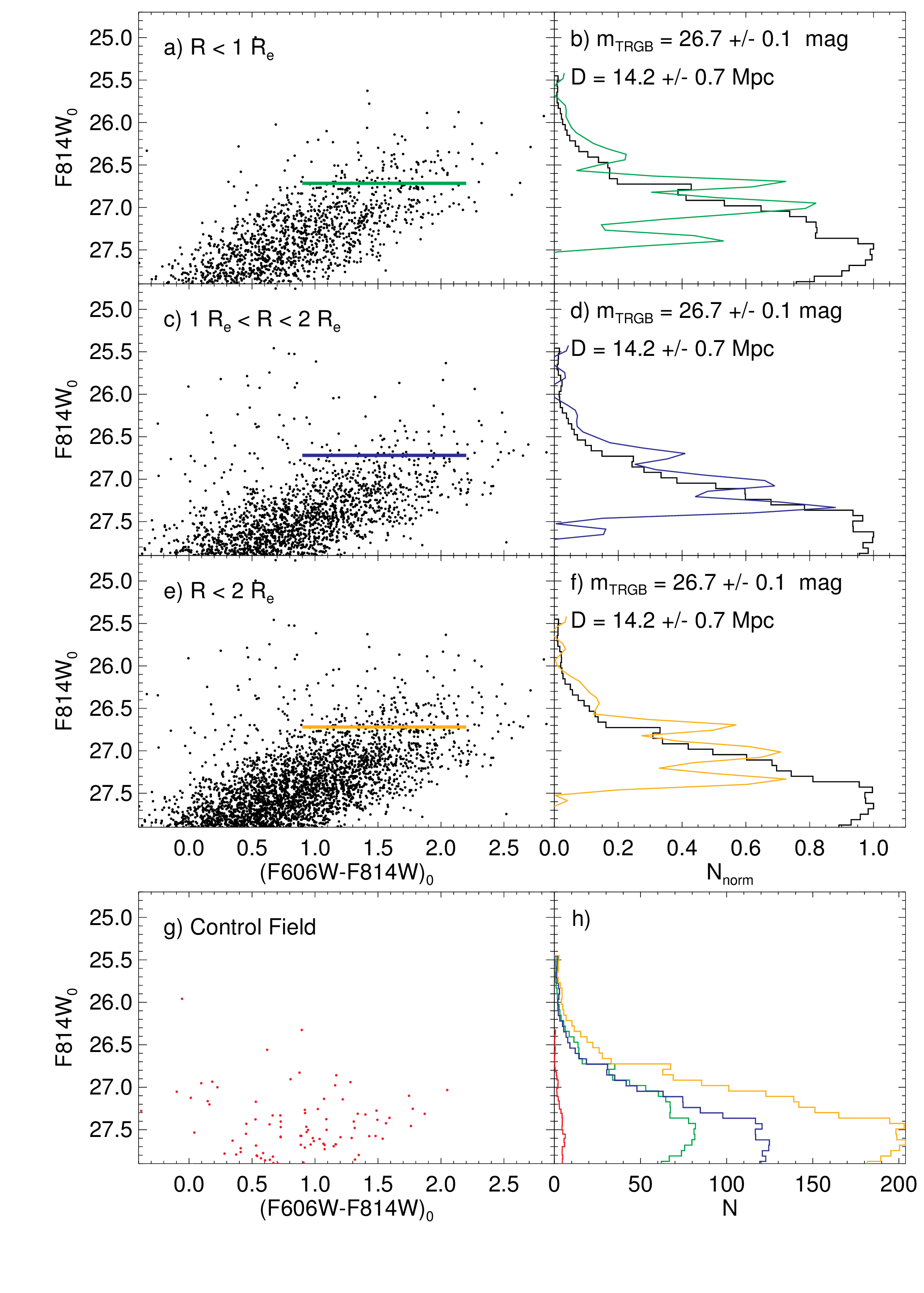}
      \vspace{-1cm}
  \caption{{\em Left panels a),c),e)-} De-reddened CMD of NGC1052-DF4 stars as a function of radius: within 1 R$_e$ (R$<$16.5 arcsec, panel a), between 1 R$_e$ and 2  R$_e$ (panel c), and within 2  R$_e$ (panel e). Panel g) presents the CMDs of a Control Field of similar area located at $\sim$7.5  R$_e$ from the center of NGC1052-DF4.   {\em Right panels b),d),f)} Luminosity functions (all normalized to peak at one; black histograms) and  filter responses (coloured lines). These were used to identify the location of the TRGB, which is marked by the horizontal lines in the left panels.  {\em Panel h)} presents the superposition of the three LFs (using the same color used to display the corresponding filter curve) compared to the Control Field one (in red). While
  the jump in the LF due to TRGB clearly happens at very similar magnitude for the first three curves,
  the Control Field presents a significantly different distribution, with the LF peaking $\approx$ 1 mag
  fainter.}
     \label{cmds}
  \end{figure*}

    \begin{figure}
  \includegraphics[width=\columnwidth]{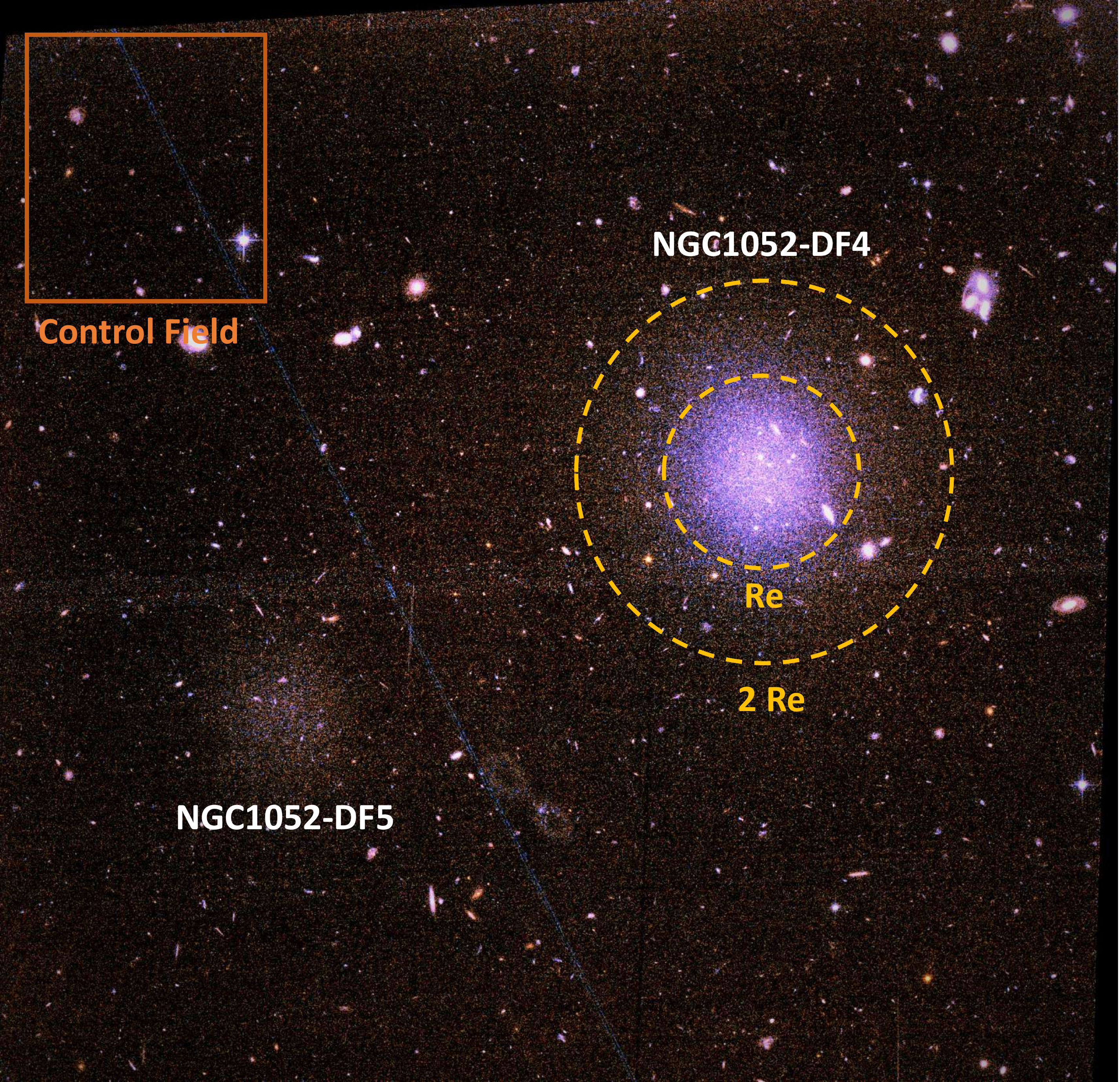}

  \caption{Colour composite image of the ACS data used in this work showing the location of the galaxies NGC1052-DF4 and NGC1052-DF5. The regions enclosing 1 and 2 R$_e$ of DF4 are shown using dashed yellow circles. The location on the Control Field is indicated with an orange square.}

     \label{fig:df4df5}
  \end{figure}

The resulting modified and de-reddened CMDs as a function of radius are shown in the left panels of Fig. \ref{cmds}. The location of the areas used to explore the CMDs are displayed in Fig. \ref{fig:df4df5}. Fig. \ref{cmds} shows, from top to bottom, the CMD of sources within 1 R$_e$ \citep[R$_e$=16.5 arcsec;][panel {\em a)}]{2018ApJ...868...96C}, between 1 and 
2 R$_e$ (panel {\em c)}), and within 2 R$_e$ (panel {\em e)}). The central position of the galaxy was centred at coordinates (R.A.=02:39:15.1 ; Dec=-08:06:58.6) which corresponds to the pixel position (2868.7, 299.5) of chip 2 of the adopted reference image. The horizontal coloured lines mark the position of the TRGB, identified from the luminosity function (LF) displayed as a black histogram in the corresponding right panels. The three LFs present a sudden jump at magnitude \textsl{F814W$_{0, Vega}$} $\approx$ 26.7 mag. The statistical significance of this feature (assuming a Poissonian distribution) is 3.2$\sigma$ (99.86\%), 2.2$\sigma$ (97.2\%) and 4.0$\sigma$ (99.99\%) (in panel {\em a)}, {\em c)} and {\em e)} respectively). The LF function was convolved with a Sobel kernel \citep{1996ApJ...461..713S,1997ApJ...478...49S} K=[-2,-1,0,1,2]. The position of the apparent magnitude of the TRGB (labelled in the right panels) was derived identifying the peak corresponding to the largest jump in the LF.

The distance modulus was derived with the \citet{2007ApJ...661..815R} zero points, obtaining $(m-M)_0$ = 30.76 $\pm$ 0.10 mag, corresponding to a distance D = 14.2 $\pm$ 0.7 Mpc\footnote{Note that the distance to DF4 is compatible with the lower limit TRGB distance ($>$9.7 Mpc) estimated in \citet{2018ApJ...868...96C}. It is worth mentioning that we used a different culling criteria than \citet{2018ApJ...868...96C} to extract our photometry, as ours is based on \texttt{DAOPHOT} while theirs was based on \texttt{DOLPHOT}.}. The final error budget includes the uncertainty of the calibration relation by \citet{2007ApJ...661..815R} and the error on the determination of the TRGB position. We also checked how the location of the TRGB would change, if instead of using $|sha| >$ 0.1, we would have used the less conservative $|sha| >$ 0.5. In that case, the location of the TRGB would be brighter by 0.1 mag, making the galaxy closer by 0.7 Mpc (i.e. D=13.5 $\pm$ 0.7 Mpc).

The excellent agreement between the distance determination in the innermost region (R $<$ R$_e$, panel {\em a)})
and in a more external annulus (1R$_e$ $<$ R $<$ 2R$_e$, panel {\em c)}), strongly suggests that the contamination
by blends (if any) is marginally affecting our distance estimation. Panel {\em g)} of Fig. \ref{cmds} presents the CMD of a Control
Field with an area of 50$\times$50 arcsec$^2$
located at the coordinates ($R.A.$,$Dec$) = (02:39:19.2; -08:08:46.7). These coordinates correspond to the pixel position (500,1500) of chip 2 of the adopted reference image. The center
of the Control Field is about 7.5 R$_e$ away from the center of NGC1052-DF4. The area of the Control Field (2500 arcsec$^2$) is chosen to represent an area similar to the one used for exploring the CMD of the galaxy (855 arcsec$^2$ within 1 R$_e$ and 2565 arcsec$^2$ between 1 and 2 R$_e$). The negligible, if any, presence of stars belonging to NGC1052-DF4 in the location of the Control Field results in a completely different distribution of sources in its CMD, as well as in the LF peaking at significantly fainter magnitude (panel {\em h)}, red line).

\begin{figure*}
  \includegraphics[width=\textwidth]{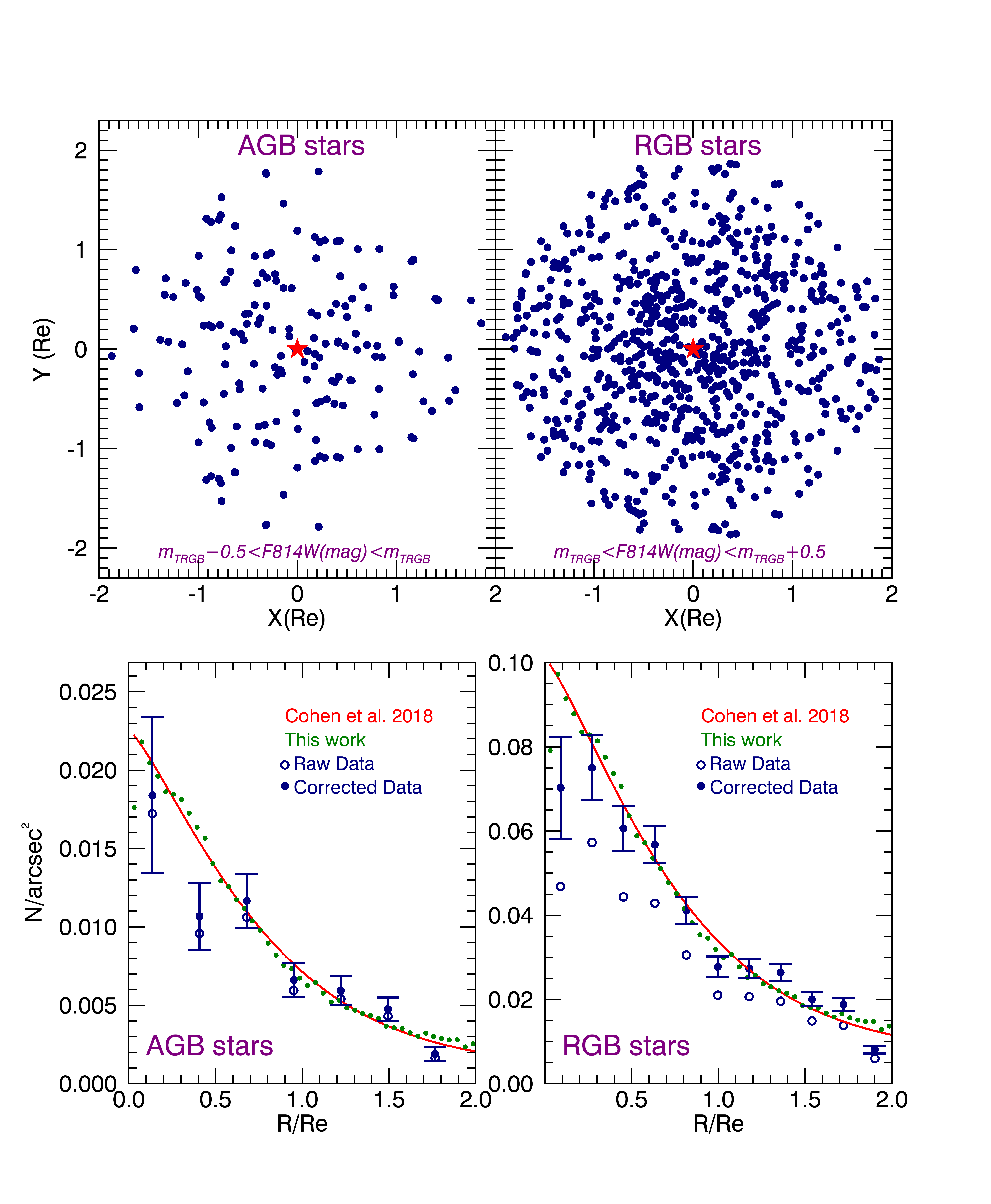}
      \vspace{-1cm}
  \caption{Spatial distribution and radial profiles of the stars detected above (AGBs) and below (RGBs) the TRGB of NGC1052-DF4. {\em Upper-panels:} Spatial distribution around the central position of the galaxy (red cross). The distance is indicated in R$_e$ units. Note how the stars found tend to group around the central part of NGC1052-DF4 as expected.  {\em Lower-panels:} Number density profiles of the stars detected around the TRGB. Open blue symbols correspond to the number density of the stars detected without applying any completeness correction, while solid blue symbols are the number of stars estimated after the completeness correction. Overplotted to the radial profiles of the stars are the surface brightness profile of the galaxy obtained using the \texttt{IRAF ELLIPSE} code (green points) and the S\'ersic model fit to the surface brightness distribution provided in \citet{2018ApJ...868...96C} (red line).}
     \label{profiles}
  \end{figure*}

To further support that the detection of the TRGB is not being significantly affected by noise peaks or possible blends, we have conducted the following diagnostic. We have explored the number of detected stars above and below the TRGB as a function of the radial distance. This is done in Fig. \ref{profiles}. As we are using only chip 2 in our analysis, to fill the distribution of stars below -0.9R$_e$ we have mirrored the stars above +0.9R$_e$. We have divided the detected stars in two groups: AGBs (stars that are brighter than the TRGB) and RGBs (stars that are fainter than the TRGB). Both groups expand 0.5 mag above and below the magnitude of the TRGB. The upper panels of Fig. \ref{profiles} show the spatial location of the AGBs and RGBs around the central position of the galaxy. The lower panels of such figure show the number density of the detected stars as a function of radius. The figure clearly indicates that both AGBs and RGBs tend to cluster around the central part of the galaxy. More importantly, the number density of the stars declines following exactly the same trend as the surface brightness distribution of the object. In Fig. \ref{profiles} we draw both the number density of stars as observed directly (open symbols) and corrected by the incompleteness of the data (filled symbols). Incompleteness simulations were done by injecting a large number of artificial stars (200000) of different magnitude across the image. For comparison, we overplotted the surface brightness distribution obtained on the F814W using the \texttt{IRAF ELLIPSE} routine (green dotted points) and the S\'ersic model fitting solution to the surface brightness distribution obtained by \citet{2018ApJ...868...96C} (red line). As it can be seen the agreement is excellent.

\section{Distances to the brightest galaxies in the field-of-view of NGC1052}

Unfortunately, none of the brightest galaxies in the field-of-view of NGC1052 (i.e. R$<$30 arcmin centered on this object) has the distance estimated using the TRGB method. This implies that their distances have been measured using less reliable techniques as the SBF or the Tully-Fisher relation.

\subsubsection{The distance to NGC1042}

\begin{figure}
  \includegraphics[width=\columnwidth]{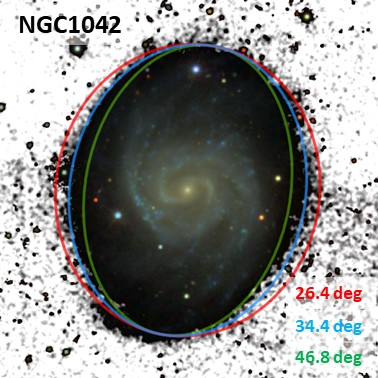}
    \includegraphics[width=\columnwidth]{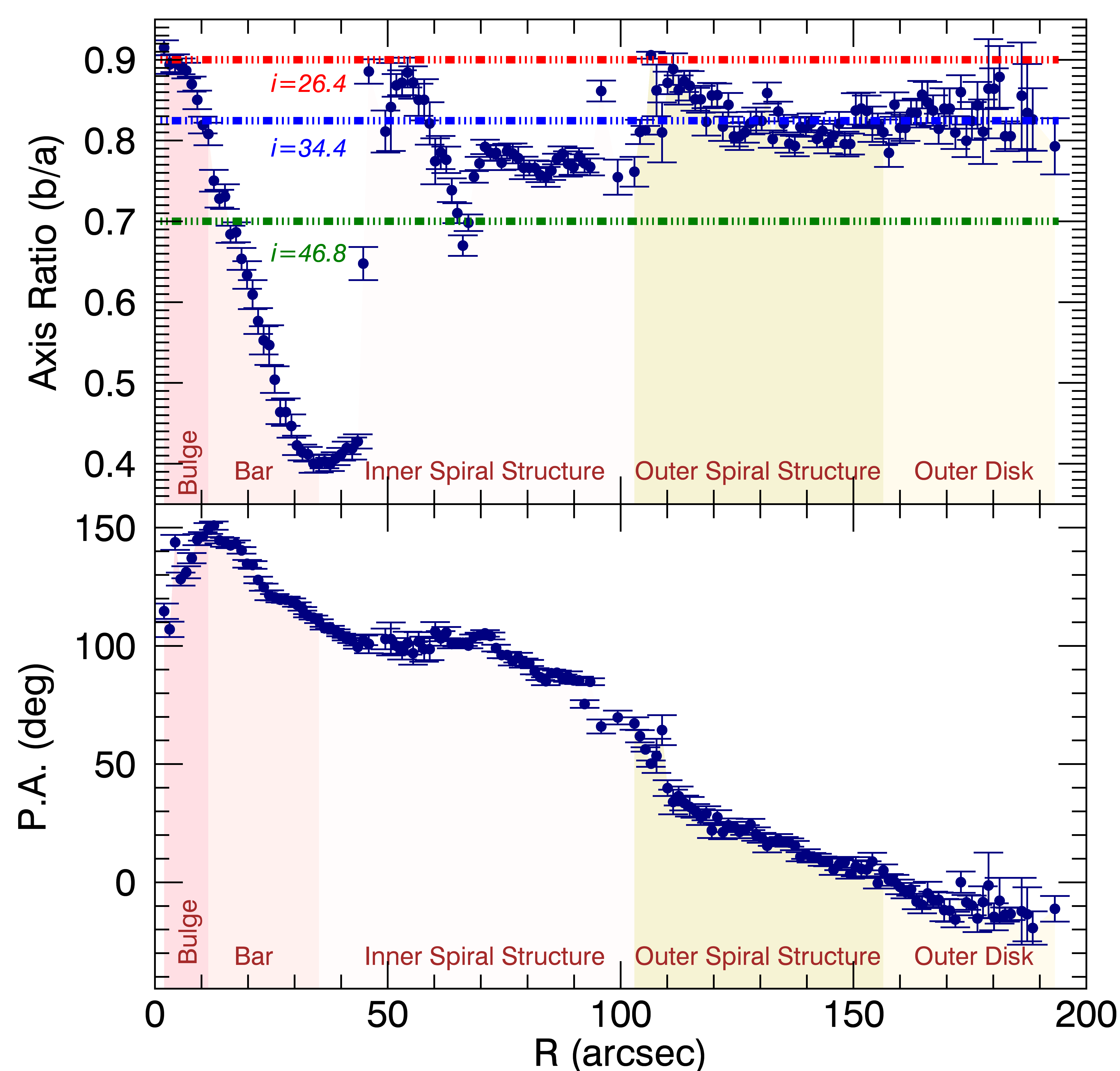}
  \caption{The inclination of NGC1042. Having an accurate estimation of the inclination of NGC1042 is key to get a reliable measurement of the distance to the galaxy through the Tully-Fisher relation. The top panel shows the SDSS image of NGC1042 with three elliptical contours overplotted with different axis ratio. The image is a color composite of the SDSS filters \textsl{g}, \textsl{r} and \textsl{i}, being the black and white background the sum of all the filters. The blue line corresponds to the inclination 34.4 deg we measure in this work. Together with such axis ratio, we show the corresponding ellipse representing the expected isophotal shape for an inclination of 26.4 deg (b/a=0.9; red line) and 46.8 deg (b/a=0.7; green line). For all the cases we have assumed q$_0$=0.2 (see text for details). The lower panel shows the axis ratio and position angle profiles of NGC1042 obtained with the \texttt{IRAF ELLIPSE} package.
 Inclinations lower than 30 degrees are not favoured by the shape of the external isophotes of the galaxy. A distance of 19 Mpc would require b/a=0.898 (i.e. 26.7 deg) which is ruled out by the observations. }

     \label{fig:ngc1042}
  \end{figure}

The distance to NGC1042 has been measured using different methods. On one hand, there have been attempts to measure the distance using the reconstruction of the Large Scale Structure along the line of sight of the object. The most accurate measurement, which made a full model of the peculiar velocity to the Galaxy within $80\;\mathrm{Mpc}$ and also corrects the peculiar velocity of NGC 1042 provides a distance of 13.2 Mpc \citep[parameter d$_k^c$ in Table 5 of ][]{2007A&A...465...71T}. On the other hand, the distance to the galaxy has been explored by mean of the Tully-Fisher relation. This last approach is the one producing a larger scatter on the values to the system, ranging from $\sim$4 \citep{2008ApJ...676..184T} to $\sim$8 \citep{1992ApJS...80..479T} Mpc. The main reason for this uncertainty is the lack of an accurate estimation of the inclination of the galaxy. In fact, values as low as 37$\degree$ have been measured \citep[][ based on 2MASS photometry]{2016ApJ...823...85L} but also larger values have been also provided \citep[57$\degree$;][]{2008ApJ...676..184T}. The reason for such discrepancies is that the axis ratio and position angle of the galaxy change along the radial distance due to the presence of a bar and the spiral arms of the object (see  Fig. \ref{fig:ngc1042}). To overcome this difficulty and have an accurate estimation of the inclination of NGC1042, it is necessary to measure the ellipticity of the isophotes of the galaxy in its outer parts, a region where the internal structure of the galaxy is not playing any role.

In order to provide an accurate estimation of the Tully-Fisher distance to this object we have measured the inclination of the galaxy at its periphery (155$\arcsec$$<$R$<$195$\arcsec$). At such a radial distance the galaxy is not affected by the internal complex structure. In fact, both the axis ratio and the position angle remain pretty stable at such location (see  Fig. \ref{fig:ngc1042}). We have followed the recipe provided by \citet{2012ApJ...749...78T} to get the distance to the object. This recipe requires three parameters: a) a measure of the rotation of the galaxy by using the 21 cm HI spectral line. For that, we use the value provided by \citet{1986AJ.....91..791V} in their Fig. 14 (i.e. V$_{max}$-V$_{min}$$\sim$145 ($\sin i$)$^{-1}$ km s$^{-1}$); b) the apparent luminosity of the galaxy given in the \textsl{I} band; and c) the inclination of the galaxy obtained from the ellipticity of the photometric image. 

As a proxy for the \textsl{I} band, we use the SDSS \textsl{i}-band. We have retrieved the \textsl{g}, \textsl{r} and \textsl{i}-band from the SDSS survey. This data is deep enough to provide accurate photometry down to 26.5 mag/arcsec$^2$ \citep[\textsl{r}-band;][]{2006A&A...454..759P}. We use the axis ratio of the isophotes in the \textsl{i}-band to estimate the axis ratio of the galaxy. In particular, we use the average value of the axis ratio of the isophotes in the radial range 155$\arcsec$$<$R$<$195$\arcsec$. The mean axis ratio is 0.832$\pm$0.004. In Fig. \ref{fig:ngc1042}, we plot the shape of an elliptical contour with such axis ratio at a radial distance of 190$\arcsec$. The contour represents very well the outer part of the object. Once we have the axis ratio, we estimate the inclination of the galaxy using the following expression $\cos i=[((b/a)^2-q_0^2)/(1-q_0^2)]^{1/2}$ with q$_{0}$=0.2  \citep{2012ApJ...749...78T}. This results in $\cos i$=0.824 (i.e. i=34.4$\degree$).

In order to measure the \textsl{I} band magnitude of the galaxy, we apply the following transformation from the SDSS photometry to the Cousin system:  \textsl{I}=\textsl{r}-1.2444$\times$(\textsl{r}-\textsl{i})-0.3820 (\url{http://classic.sdss.org/dr5/algorithms/sdssUBVRITransform.html#vega_sun_colors}; Lupton 2005). The SDSS apparent (AB) magnitudes we measure correspond to the integrated magnitudes down to the i-band 26 mag/arcsec$^2$ isophote. These are: 11.24$\pm$0.05 mag (\textsl{g}-band), 10.80$\pm$0.05 mag (\textsl{r}-band), 10.48$\pm$0.05 mag (\textsl{i}-band). This results in I=10.02$\pm$0.07 mag.  We can compare this value with the one provided for this galaxy in the Extragalactic Distance Database (EDD) catalogue \citep[I=10.19 mag; column 23 in Table 1 of ][]{2008ApJ...676..184T}. Using the following Galactic extinction correction for the SDSS filters (0.065 and 0.049 mag; \textsl{r} and \textsl{i}-bands respectively), the observed Vega \textsl{I} band magnitude after correcting the Galactic extinction is 9.97$\pm$0.07 mag. To correct for the effect of the internal extinction of NGC1042 we use the following expression: A$_{i}^{'}$=$\gamma_I$$\log$(a/b) where a/b is the major to minor axis ratio and $\gamma_I$ is given by \citet{2012ApJ...749...78T}: 

\begin{equation}
\gamma_I=0.92+1.63(\log (V_{max}-V_{min})-2.5)   
\end{equation}
obtaining $\gamma_I$=0.725, so A$_{i}^{'}$=0.068 mag. The result of applying the internal and Galactic extinction correction is \textsl{I}=9.91 mag for NGC1042.

Finally, we measure the distance modulus to NGC1042 using the following expression \citep{2012ApJ...749...78T}:

\begin{equation}
    \mu=I-M_I=I+21.39+8.81(\log (V_{max}-V_{min})-2.5)
\end{equation}
which results in $\mu$=30.24$\pm$0.41 and a distance of 12.6$\pm$2.3 Mpc. Using the I magnitude provided in the EDD catalogue  would give 13.5$\pm$2.6 Mpc. In order to have the galaxy located at 19 Mpc, the inclination of the object would need to be as low as 26.7 deg (b/a=0.898)\footnote{Using the EDD I-band magnitude, I=10.19 mag, the inclination would need to be 28.0 deg (b/a=0.888), again far away from the observations.}, which is incompatible with the shape of its external isophotes (see Fig. \ref{fig:ngc1042}). Moreover, at all radial distances the inclination of the galaxies is not compatible with being close to a face-on projection as suggested by \citet{2019RNAAS...3b..29V}. It is finally worth noting that the Tully-Fisher distance of NGC1042 is compatible, within the error bars, with the distance obtained from the Large Scale Structure reconstruction \citep[13.2 Mpc;][]{2007A&A...465...71T}.

\subsubsection{The distance to NGC1052 and NGC1047}

The distance to NGC1052 has been estimated using the surface brightness fluctuations (SBF) technique both in the \textsl{I} band \citep[19.4$\pm$2 Mpc;][]{2001ApJ...546..681T} and in the \textsl{F160W} band \citep[18$\pm$2 Mpc;][]{2003ApJ...583..712J}. The two results agree on a distance compatible with $\sim$19 Mpc\footnote{The Large Scale Structure reconstruction done by \citet{2007A&A...465...71T} gives also a compatible distance of 17.4 Mpc for this object.}. Interestingly, NGC1052 is interacting with another galaxy in the field-of-view: NGC1047. This is clearly shown in the Dragonfly deep image of the group shown in Fig. 4 of \citet{2019ApJ...874L...5V} and the deep image by the Heron telescope provided by Fig. 1 of \citet{2019A&A...624L...6M}. This interaction is also very likely  responsible for the perturbed distribution of HI around NGC1052 found in \citet{1986AJ.....91..791V}. In fact, this idea fits very well with the tail of HI gas of NGC1052 directly pointing to NGC1047 \citep[see Fig. 5 of][]{1986AJ.....91..791V}. It is worth noting that distance quoted in the literature for NGC1047 of 4.6 Mpc \citep{1984A&AS...56..381B} is clearly incompatible with the ongoing interaction between this object and NGC1052.

.

\subsubsection{The distance to NGC1035}

Another bright galaxy in the FOV of NGC1052 is the spiral NGC1035. This galaxy is particularly interesting due to its vicinity to NGC1052-DF4. The most recent Tully-Fisher distance determination for this galaxy has been obtained by \citet{2014MNRAS.444..527S}. These authors quote a selection-bias-corrected distance estimate of 14$\pm$2.9 Mpc for NGC1035. This makes the galaxy distance compatible with NGC1042.

\begin{figure*}
	\includegraphics[width=\textwidth]{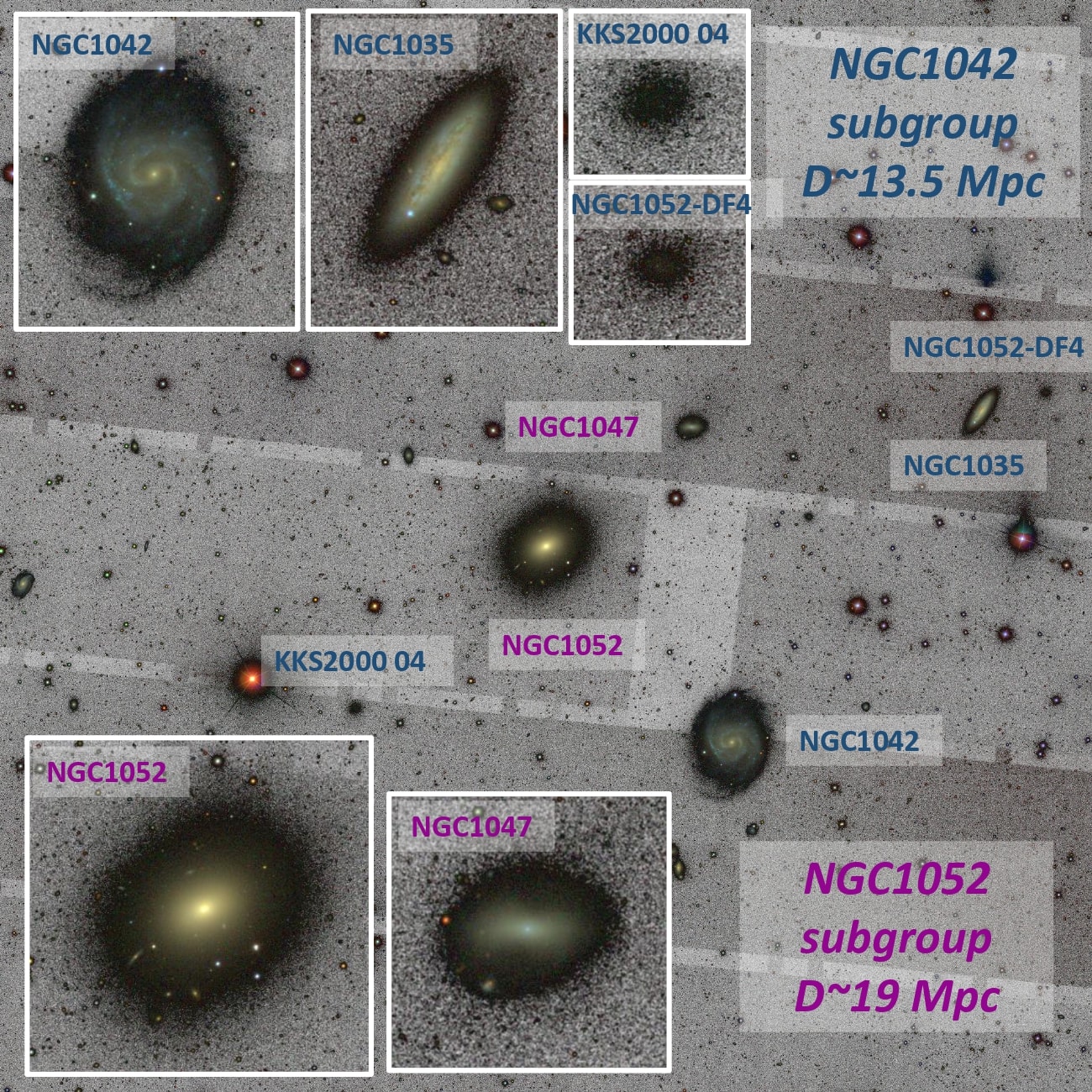}
    \caption{Distribution of the distances of the galaxies in the field-of-view of NGC1052. The galaxies are distributed around two different distances, one at 13.5 Mpc containing NGC1042 and NGC1035 (and the dwarfs [KKS 2000] 04 and NGC1052-DF4), and another group at 19 Mpc containing NGC1052 and NGC1047. The image is a color composite of the SDSS filters \textsl{g}, \textsl{r} and \textsl{i}, being the black and white background the sum of all the filters.}
    \label{fig:distances}
\end{figure*}

Finally, there is another bright galaxy located at 28.1$\arcmin$ from NGC1052: NGC1069. This object, however, has a heliocentric radial velocity of $\sim$9400 km s$^{-1}$ which makes it incompatible with being physically associated either to NGC1042 or NGC1052. The distance estimations for the galaxies in the field-of-view of NGC1052 are summarised in Table \ref{table:distances} and Fig. \ref{fig:distances}.

\begin{table*}
\centering
\caption{Distances to the galaxies in the FOV of NGC1052. The foreground extinction in the HST bands \textsl{F606W} and \textsl{F814W} are indicated for reference. The velocity $v_{hel}$ corresponds to the mean heliocentric radial velocity. In order to make a sensible comparison of the velocities of these galaxies with the Hubble Flow, we include also the velocities with respect to the Cosmic Microwave Background (CMB) $v_{CMB}$ which have been estimated subtracting 217 km/s to $v_{hel}$ taking into account the location on the sky of these galaxies and the solar motion of 370.06 km/s towards the direction defined by Galactic coordinates (263.914°, +48.2646°). See HYPERLEDA \citep{2014A&A...570A..13M} for further details.}
\begin{tabular}{ccccccccc} 
\hline
\label{table:distances}
Name & RA      &  Dec     & Distance & A$_{F606W}$ & A$_{F814W}$ & $v_{hel}$ & $v_{CMB}$ & Method  \\
     & (J2000) & (J2000)  & (Mpc)    & (mag) & (mag)  & (km/s) & (km/s) & \\
\hline
& & & Diffuse galaxies & & &  \\
\hline
[KKS2000]04 & 02:41:46.8  & -08:24:09.3  & 13.0$\pm$0.4 & 0.060 & 0.037 & 1793$\pm$2 & 1576$\pm$2 & TRGB \\ 
NGC1052-DF4 & 02:39:15.1  & -08:06:58.6  & 14.2$\pm$0.7 & 0.062 & 0.038 & 1445$\pm$4 & 1228$\pm$4 & TRGB \\ 
\hline
& & & Bright galaxies & &  \\
\hline
NGC1035 & 02:39:29.1 & -08:07:58.6 & 14$\pm$3  & 0.062 & 0.038 & 1262$\pm$16 & 1045$\pm$16 & Tully-Fisher \\
NGC1042 & 02:40:24.0 & -08:26:01.0 & 12.6$\pm$2.3 & 0.071 & 0.044 & 1371$\pm$1 & 1155$\pm$1 & Tully-Fisher\\
NGC1047 & 02:40:32.8 & -08:08:51.4 & 19$\pm$2 & 0.063 & 0.039 & 1415$\pm$15 & 1198$\pm$15 & Physical association with NGC1052 \\
NGC1052 & 02:41:04.8 & -08:15:20.8 & 19$\pm$2 & 0.066 & 0.041 & 1484$\pm$6 & 1268$\pm$6 & SBF \\
\hline
\end{tabular}
\end{table*}

\section{Discussion}

The results shown in this work reinforce the idea that the galaxies in the line-of-sight of NGC1052 are grouped into two structures. We argue that the closest structure is formed by NGC1042 and NGC1035
and contains the dwarf galaxies: [KKS2000 04] and NGC1052-DF4. This group of galaxies is located at $\sim$13.5 Mpc. In projection, we then have a second group, whose principal galaxy is NGC1052 and  also contains NGC1047. This group is  located at $\sim$19 Mpc. Despite being placed at different distances, the two groups share similar heliocentric velocities. When these velocities are transformed to the CMB rest-frame and therefore a comparison with the Hubble Flow can be done, we find the following. The average velocity of the two most massive galaxies located at $\sim$13.5 Mpc (NGC1042 and NGC1035) is 1100 km/s (i.e. the velocity of this group deviates from the expected Hubble Flow at that distance by $\sim$+150 km/s), while the average velocity of the two massive galaxies located at 19 Mpc (NGC1047 and NGC1052) is 1233 km/s (i.e. the departure from the Hubble Flow is $\sim$-100 km/s). In both cases, the absolute deviation from the Hubble Flow is very similar and not very large. We refer the reader to \citet{2019MNRAS.486.1192T} for an in-depth analysis of the largest discrepant galaxy [KKS2000 04]. In short, the values of the velocities of  the galaxies in these two groups are not at all unexpected considering the typical peculiar velocities, in comparison to the Hubble flow, observed among the  nearby galaxies \citep[see a  discussion about this in Section 5 of ][]{2019MNRAS.486.1192T}. A distance of 13.5 Mpc would also make the globular clusters of DF4 to have a less extreme nature. Their intrinsic brightness would be fainter and their sizes would be smaller, going from a median size of 4.1 pc \citep{2019ApJ...874L...5V} to 2.8 pc, which is in better agreement with the median size of both globular clusters in the  Milky Way (2.9$^{+0.1}_{-0.2}$ pc) and dwarf galaxies (3.2$^{+0.1}_{-0.2}$ pc) \citep[see e.g.][]{2019MNRAS.486.1192T}. 

Finally, it is worth mentioning how the effect of the distance impact on the claim of the dark matter content of NGC1052-DF4. \citet{2019ApJ...874L...5V} infer a distance for this galaxy of 20 Mpc. With the TRGB distance measured here, 14.2$\pm$0.7 Mpc,  the dynamical mass of the galaxy (assuming an isothermal dark matter halo) decreases by a factor of 20/14.2  (i.e. $\sim$1.4) while its stellar mass decreases by a larger factor of 10$^{0.4\times(31.505-30.761)}$ (i.e. $\sim$2). \citet{2019ApJ...874L...5V} also assumes a (M/L)$_{stars,V}$=2.0$\pm$0.5$\Upsilon$$_{\odot}$, while we find, for these types of diffuse galaxies, based on their Spectral Energy Distribution (from UV to IR) and assuming a Chabrier Initial Mass Function (IMF), a value of (M/L)$_{stars,V}$=1.07$^{+0.80}_{-0.54}$ $\Upsilon$$_\odot$ \citep[][]{2019MNRAS.486.1192T}, which decreases by another factor of two the stellar mass content\footnote{\citet{2019MNRAS.486.5670R} have recently measured spectroscopically the age and metallicity of [KKS2000 04]: 8.7 Gyr and [M/H]=-1.18. With these estimations and assuming a Chabrier IMF (M/L)$_{stars,V}$=1.56 $ \Upsilon_\odot$.}. The combination of these two factors would reduce the total stellar mass of NGC1052-DF4 by a factor of $\sim$3, alleviating significantly the absence of dark matter previously reported. To end this discussion, it is also worth mentioning the strong relevance the assumption on the shape of the dark matter halo has for measuring the dark matter content of these galaxies. In \citet{2019MNRAS.486.1192T}, we showed that the total dark matter halo that can accommodate the dynamics of KKS2000 04 changes by a factor of $\sim$3 from 10$^{9.1}$ M$_{\odot}$ (using a NFW halo) to 10$^{9.6}$ M$_{\odot}$ (using a  Burkert cored halo with core radius of 4 kpc). In short, the larger the core of the dark matter halo the larger is the amount of dark matter compatible with the observed dynamics. There is growing evidence (both theoretical and observational) that the shape of the dark matter halo of the ultra diffuse galaxies is probably dominated by an extended dark matter core \citep[see for instance the works by][]{2014MNRAS.441.2986D,2019arXiv190404838V}. All in all, the proposition that both KKS2000 04 and NGC1052-DF4 galaxies are ``missing dark matter" is still far from being placed on sure footing.

\acknowledgments

We thank the referee for his/her careful reading of the manuscript which help to improve the quality of the analysis presented in this work. We thank Nushkia Chamba for her help on the analysis of the properties of NGC1042. Mireia Montes, Javier Rom\'an and Mike Beasley provided a lot of interesting comments.
I.T. acknowledges financial support from the European Union's Horizon 2020 research and innovation programme under Marie Sklodowska-Curie grant agreement No 721463 to the SUNDIAL ITN network. We also acknowledge support from the Fundaci\'on BBVA under its 2017 programme of assistance to scientific research groups, for the project "Using machine-learning techniques to drag galaxies from the noise in deep imaging". This research has been partly supported by the Spanish Ministry of Economy and Competitiveness (MINECO) under grants AYA2016-77237-C3-1-P and AYA2017-89076-P.


\begin{thebibliography}{}

\bibitem[Aloisi et al.(2007)]{2007ApJ...667L.151A} Aloisi, A., Clementini, G., Tosi, M., et al.\ 2007, \apjl, 667, L151 

\bibitem[Blakeslee et al.(2010)]{2010ApJ...724..657B} Blakeslee, J.~P., Cantiello, M., Mei, S., et al.\ 2010, \apj, 724, 657 

\bibitem[Blakeslee \& Cantiello(2018)]{2018RNAAS...2c.146B} Blakeslee, J.~P., \& Cantiello, M.\ 2018, Research Notes of the American Astronomical Society, 2, 146 

\bibitem[Bottinelli et al.(1984)]{1984A&AS...56..381B} Bottinelli, L., Gouguenheim, L., Paturel, G., \& de Vaucouleurs, G.\ 1984, \aaps, 56, 381 

\bibitem[Cantiello et al.(2018)]{2018ApJ...856..126C} Cantiello, M., Blakeslee, J.~P., Ferrarese, L., et al.\ 2018, \apj, 856, 126 

\bibitem[Carlsten et al.(2019)]{2019arXiv190107575C} Carlsten, S., Beaton, R., Greco, J., \& Greene, J.\ 2019, arXiv:1901.07575 

\bibitem[Cohen et al.(2018)]{2018ApJ...868...96C} Cohen, Y., van Dokkum, P., Danieli, S., et al.\ 2018, \apj, 868, 96

\bibitem[Di Cintio et al.(2014)]{2014MNRAS.441.2986D} Di Cintio, A., Brook, C.~B., Dutton, A.~A., et al.\ 2014, \mnras, 441, 2986 

\bibitem[Lee et al.(1993)]{1993ApJ...417..553L} Lee, M.~G., Freedman, W.~L., \& Madore, B.~F.\ 1993, \apj, 417, 553 

\bibitem[Luo et al.(2016)]{2016ApJ...823...85L} Luo, R., Hao, L., Blanc, G.~A., et al.\ 2016, \apj, 823, 85 

\bibitem[Jensen et al.(2003)]{2003ApJ...583..712J} Jensen, J.~B., Tonry, J.~L., Barris, B.~J., et al.\ 2003, \apj, 583, 712 

 \bibitem[Makarov et al.(2014)]{2014A&A...570A..13M} Makarov D., Prugniel P., Terekhova N., Courtois H., \& Vauglin I.\ 2014, \aap, 570, A13 

\bibitem[McQuinn et al.(2017)]{mcquinn17a} McQuinn, K.~B.~W., Skillman, E.~D., Dolphin, A.~E., Berg, D., \& Kennicutt, R.\ 2017, \aj, 154, 51

\bibitem[Monelli et al.(2010)]{2010ApJ...720.1225M} Monelli, M., Hidalgo, S.~L., Stetson, P.~B., et al.\ 2010, \apj, 720, 1225


\bibitem[M{\"u}ller et al.(2019)]{2019A&A...624L...6M} M{\"u}ller, O., Rich, R.~M., Rom{\'a}n, J., et al.\ 2019, \aap, 624, L6

\bibitem[Pohlen \& Trujillo(2006)]{2006A&A...454..759P} Pohlen, M., \& Trujillo, I.\ 2006, \aap, 454, 759 

\bibitem[Rizzi et al.(2007)]{2007ApJ...661..815R} Rizzi, L., Tully, R.~B., Makarov, D., et al.\ 2007, \apj, 661, 815 

\bibitem[Ruiz-Lara et al.(2019)]{2019MNRAS.486.5670R} Ruiz-Lara, T., Trujillo, I., Beasley, M.~A., et al.\ 2019, \mnras, 486, 5670 

\bibitem[Sakai et al.(1996)]{1996ApJ...461..713S} Sakai, S., Madore, B.~F., \& Freedman, W.~L.\ 1996, \apj, 461, 713 

\bibitem[Sakai et al.(1997)]{1997ApJ...478...49S} Sakai, S., Madore, B.~F., Freedman, W.~L., et al.\ 1997, \apj, 478, 49 

\bibitem[Salaris \& Cassisi(1997)]{1997MNRAS.289..406S} Salaris, M., \& Cassisi, S.\ 1997, \mnras, 289, 406 

\bibitem[Skillman et al.(2017)]{2017ApJ...837..102S} Skillman, E.~D., Monelli, M., Weisz, D.~R., et al.\ 2017, \apj, 837, 102 

\bibitem[Sorce et al.(2014)]{2014MNRAS.444..527S} Sorce, J.~G., Tully, R.~B., Courtois, H.~M., et al.\ 2014, \mnras, 444, 527 


\bibitem[Stetson(1987)]{stetson87} Stetson, P.~B.\ 1987, \pasp, 99, 191 

\bibitem[Stetson(1994)]{stetson94} Stetson, P.~B.\ 1994, \pasp, 106, 250

\bibitem[Theureau et al.(2007)]{2007A&A...465...71T} Theureau, G., Hanski, M.~O., Coudreau, N., Hallet, N., \& Martin, J.-M.\ 2007, \aap, 465, 71

\bibitem[Tonry et al.(2001)]{2001ApJ...546..681T} Tonry, J.~L., Dressler, A., Blakeslee, J.~P., et al.\ 2001, \apj, 546, 681 

\bibitem[Trujillo et al.(2019)]{2019MNRAS.486.1192T} Trujillo, I., Beasley, M.~A., Borlaff, A., et al.\ 2019, \mnras, 486, 1192 

\bibitem[Tully et al.(1992)]{1992ApJS...80..479T} Tully, R.~B., Shaya, E.~J., \& Pierce, M.~J.\ 1992, \apjs, 80, 479 

\bibitem[Tully et al.(2008)]{2008ApJ...676..184T} Tully, R.~B., Shaya, E.~J., Karachentsev, I.~D., et al.\ 2008, \apj, 676, 184 

\bibitem[Tully \& Courtois(2012)]{2012ApJ...749...78T} Tully, R.~B., \& Courtois, H.~M.\ 2012, \apj, 749, 78 

\bibitem[Turri et al.(2017)]{2017AJ....153..199T} Turri, P., McConnachie, A.~W., Stetson, P.~B., et al.\ 2017, \aj, 153, 199 

\bibitem[van Dokkum et al.(2018a)]{2018Natur.555..629V} van Dokkum, P., Danieli, S., Cohen, Y., et al.\ 2018a, \nat, 555, 629 

\bibitem[van Dokkum et al.(2019a)]{2019ApJ...874L...5V} van Dokkum, P., Danieli, S., Abraham, R., Conroy, C., \& Romanowsky, A.~J.\ 2019, \apjl, 874, L5 

\bibitem[van Dokkum et al.(2019b)]{2019RNAAS...3b..29V} van Dokkum, P., Danieli, S., Romanowsky, A., Abraham, R., \& Conroy, C.\ 2019, Research Notes of the American Astronomical Society, 3, 29 
\bibitem[van Dokkum et al.(2019c)]{2019arXiv190404838V} van Dokkum, P., Wasserman, A., Danieli, S., et al.\ 2019, arXiv:1904.04838 

\bibitem[van Gorkom et al.(1986)]{1986AJ.....91..791V} van Gorkom, J.~H., Knapp, G.~R., Raimond, E., Faber, S.~M., \& Gallagher, J.~S.\ 1986, \aj, 91, 791 

\end{thebibliography}
\end{document}